\documentstyle[12pt]{article}
\begin{document}
\large
\begin{titlepage}
\begin{center}
{\Large\bf  Compactifications of F-Theory\\ 
on Calabi-Yau Fourfolds \\}
\vspace*{15mm}
{\bf Yu. Malyuta and T. Obikhod\\}
\vspace*{10mm}
{\it Institute for Nuclear Research, National Academy of
Sciences of Ukraine\\
252022 Kiev, Ukraine\\}
{\bf e-mail: interdep@kinr.kiev.ua\\}
\vspace*{35mm}
{\bf Abstract\\}
\end{center}
\vspace*{1mm}
The enhanced gauge groups in F-theory
compactified on elliptic Calabi-Yau
fourfolds are investigated
in terms of toric geometry.
\end{titlepage}
\section {\bf Introduction}
F-theory provides a geometry/physics 
dictionary for the interpretation of solitonic states 
in terms of geometric singularities and enhanced 
gauge symmetries \cite{1.}. Recently the elegant 
algorithm for obtaining the 
enhanced gauge groups in F-theory has been 
proposed by Candelas et al. \cite{2.,3.}. This algorithm 
allows to read off the Dynkin diagram from the dual 
polyhedron that realizes the toric description  of the 
elliptic Calabi-Yau manifold. The gauge content for 
elliptic Calabi-Yau threefolds 
\begin{equation}
X_{12k+12}(1,1,2k,4k+4,6k+6) \ 
\end{equation}
($ k  = 1, \ldots,6\ $)
was obtained in \cite{4.}. 

	The purpose of the present paper is to obtain
the gauge content for elliptic Calabi-Yau fourfolds
\begin{equation}
X_{18k+18}(1,1,1,3k,6k+6,9k+9) \ 
\end{equation}
($ k  = 1, \ldots,6\ $).
\section{\bf Reflexive Polyhedra}
By means of a computer program we have computed the
dual polyhedra for the fourfolds (2). The polyhedra 
for each $ k $ have similar properties that we will 
illustrate by considering $ k=1 $ as an example. For 
this case, the points of the dual polyhedron 
$ \bigtriangledown $ are displayed in Table 1.
\newpage
\begin{center}
\begin{tabular}{|c|}  \hline
$ \bigtriangledown $
   \\   \hline \hline
$ (-1,\ 0,\ 0,\ 2,\ 3) $ \\
$ (\ 0,-1,\ 0,\ 2,\ 3) $ \\
$ (\ 0,\ 0,-1,\ 2,\ 3) $ \\
$ (\ 0,\ 0,\ 0,-1,\ 0) $ \\
$ (\ 0,\ 0,\ 0,\ 0,-1) $ \\
$ (\ 0,\ 0,\ 0,\ 0,\ 0) $ \\
$ (\ 0,\ 0,\ 0,\ 0,\ 1) $ \\
$ (\ 0,\ 0,\ 0,\ 1,\ 1) $ \\
$ (\ 0,\ 0,\ 0,\ 1,\ 2) $ \\
$ (\ 0,\ 0,\ 0,\ 2,\ 3) $ \\
$ (\ 0,\ 0,\ 1,\ 2,\ 3) $ \\
$ (\ 1,\ 1,\ 3,\ 2,\ 3) $ \\  \hline
\end{tabular}\\
\vspace*{7mm}
{\bf Table 1:} The dual polyhedron 
$ \bigtriangledown $ for $ k=1 $. 
\end{center}
\hspace*{6mm}The following observations summarize the 
structure of the polyhedron $ \bigtriangledown $\ :\\
1. Omitting the first two points and the last
point of $ \bigtriangledown $ leaves us with the dual 
polyhedron $ ^{3}\bigtriangledown $ of the $ K3 $ surface 
$ X_{12}(1,1,4,6) $\ ;\\
2. Omitting the first three points and the last
two points of $ \bigtriangledown $
 leaves us with the dual polyhedron
$ ^{2}\bigtriangledown $ of the torus $ X_{6}(1,2,3) $\ ;\\
3. The polyhedron $ ^{3}\bigtriangledown $ is divided
into a `top' and a `bottom' by the polyhedron 
$ ^{2}\bigtriangledown $ and we may write
\begin{center}
$ ^{3}\bigtriangledown $ = $ \bigtriangledown^{H}_{bot} $
$ \cup $ $ \bigtriangledown^{k=1}_{top} $,
\end{center}
where $ \bigtriangledown^{k=1}_{top} $ depends only
on $ k=1 $ while $ \bigtriangledown^{H}_{bot} $ depends
only on the enhanced gauge group $ H $.
\section{\bf Enhanced Gauge Groups}
Let us introduce special points \\
\hspace*{6cm}$ pt^{(j)}_{1}=(\ 0,\ 0,-j,\ 2,\ 3) $ , \\
\hspace*{6cm}$ pt^{(j)}_{2}=(\ 0,\ 0,-j,\ 1,\ 2) $ , \\
\hspace*{6cm}$ pt^{(j)}_{3}=(\ 0,\ 0,-j,\ 1,\ 1) $ , \\
\hspace*{6cm}$ pt^{(j)}_{4}=(\ 0,\ 0,-j,\ 0,\ 1) $ , \\
\hspace*{6cm}$ pt^{(j)}_{5}=(\ 0,\ 0,-j,\ 0,\ 0) $ , \\
\hspace*{6cm}$ pt^{(j)}_{6}=(\ 0,\ 0,-j,-1,\ 0) $  , \\
\hspace*{6cm}$ pt^{(j)}_{7}=(\ 0,\ 0,-j,\ 0,-1) $  , \\
where $ j $ is a positive integer.\\
\hspace*{6mm}We consider now the possibility of
adding to $ ^{2}\bigtriangledown $ combinations
of the points $ pt^{(j)}_{r} $ in all possible ways
such that the bottom corresponds to a reflexive
polyhedron. Tables 2 and 3 show the allowed bottoms
leading to enhanced gauge groups $ H $. Note that 
in each case the points of $ ^{2}\bigtriangledown $
are understood and $ pt^{(j)}_{r} $ implies the 
presence of $ pt^{(j-1)}_{r},\ldots,pt^{(1)}_{r} $.\\
\hspace*{6mm}Applying the analogous consideration
to tops we obtain the following gauge content
for fourfolds (2)\ :
\begin{center}
$ H \times SU(1) $ \hspace*{1cm} for \hspace*{1cm} 
$ k=1 $ ,\\
$ H \times SO(8) $ \hspace*{1cm} for \hspace*{1cm} 
$ k=2 $ , \\
$ H \times E_{6} $ \hspace*{1.8cm} for \hspace*{1cm} 
$ k=3 $ , \\
$ H \times E_{7} $ \hspace*{1.8cm} for \hspace*{1cm} 
$ k=4 $ , \\
$ H \times E_{8} $ \hspace*{1.8cm} for \hspace*{1cm} 
$ k=5 $ , \\
$ H \times E_{8} $ \hspace*{1.8cm} for \hspace*{1cm} 
$ k=6 $ . \\
\end{center}
Our computations lead to the conclusion that the types of 
singularities of elliptic fibrations (1) and (2) are
the same.
\begin{center}
\begin{tabular}{|l|l|}  \hline
$\hspace*{6mm} H $ & \hspace*{3cm} Bottom   
\\   \hline \hline
$ SU(1) $ &  \{$ pt^{(1)}_{1} $\}  \\
$ SU(2) $ & $ \{pt^{(1)}_{1},pt^{(1)}_{2}\} $ \\
$ SU(3) $ & $ \{pt^{(1)}_{1},pt^{(1)}_{2},
pt^{(1)}_{3}\} $ \\
$ G_{2} $ & $ \{pt^{(2)}_{1},pt^{(1)}_{2},
pt^{(1)}_{3}\} $    \\
$ SO(5) $ & $ \{pt^{(1)}_{1},pt^{(1)}_{2},
pt^{(1)}_{4}\} $  \\
$ SU(4) $ & $ \{pt^{(1)}_{1},pt^{(1)}_{2},pt^{(1)}_{3},
pt^{(1)}_{4}\} $   \\
$ SO(7) $ & $ \{pt^{(2)}_{1},pt^{(1)}_{2},pt^{(1)}_{3},
pt^{(1)}_{4}\} $ \\
$ Sp_{3} $ & $ \{pt^{(1)}_{1},pt^{(1)}_{2},pt^{(1)}_{4},
pt^{(1)}_{6}\} $ \\
$ SU(5) $ & $ \{pt^{(1)}_{1},pt^{(1)}_{2},pt^{(1)}_{3},
pt^{(1)}_{4},pt^{(1)}_{5}\} $ \\
$ SO(9) $ & $ \{pt^{(2)}_{1},pt^{(2)}_{2},pt^{(1)}_{3},
pt^{(1)}_{4}\} $ \\
$ F_{4} $ & $ \{pt^{(3)}_{1},pt^{(2)}_{2},pt^{(1)}_{3},
pt^{(1)}_{4}\} $ \\
$ SU(6) $ & $ \{pt^{(1)}_{1},pt^{(1)}_{2},pt^{(1)}_{3},
pt^{(1)}_{4},pt^{(1)}_{5},pt^{(1)}_{6}\} $ \\
$ SU(6)_{b} $ & $ \{pt^{(1)}_{1},pt^{(1)}_{2},
pt^{(1)}_{3},pt^{(1)}_{4},pt^{(1)}_{5},pt^{(1)}_{7}\} $ \\
$ SO(10) $ & $ \{pt^{(2)}_{1},pt^{(2)}_{2},
pt^{(1)}_{3},pt^{(1)}_{4},pt^{(1)}_{5}\} $ \\
$ SO(11) $ & $ \{pt^{(2)}_{1},pt^{(2)}_{2},pt^{(1)}_{3},
pt^{(2)}_{4},pt^{(1)}_{5}\} $ \\
$ SO(12) $ & $ \{pt^{(2)}_{1},pt^{(2)}_{2},pt^{(1)}_{3},
pt^{(2)}_{4},pt^{(1)}_{5},pt^{(1)}_{6}\} $ \\
$ E_{6} $ & $ \{pt^{(3)}_{1},pt^{(2)}_{2},pt^{(2)}_{3},
pt^{(1)}_{4},pt^{(1)}_{5}\} $ \\
$ E_{7} $ & $ \{pt^{(4)}_{1},pt^{(3)}_{2},pt^{(2)}_{3},
pt^{(2)}_{4},pt^{(1)}_{5}\} $ \\  
$ SU(6)_{c} $ & $ \{pt^{(1)}_{1},pt^{(1)}_{2},
pt^{(1)}_{3},pt^{(1)}_{4},pt^{(1)}_{5},pt^{(1)}_{6},
pt^{(1)}_{7}\} $ \\
$ SO(13) $ & $ \{pt^{(2)}_{1},pt^{(2)}_{2},
pt^{(1)}_{3},pt^{(2)}_{4},pt^{(1)}_{5},
pt^{(2)}_{6}\} $ \\
$ E_{6b} $ & $ \{pt^{(3)}_{1},pt^{(2)}_{2},
pt^{(2)}_{3},pt^{(1)}_{4},pt^{(1)}_{5},
pt^{(1)}_{7}\} $ \\
$ E_{7b} $ & $ \{pt^{(4)}_{1},pt^{(3)}_{2},
pt^{(2)}_{3},pt^{(2)}_{4},pt^{(1)}_{5},
pt^{(1)}_{6}\} $ \\
$ E_{8} $ & $ \{pt^{(6)}_{1},pt^{(4)}_{2},
pt^{(3)}_{3},pt^{(2)}_{4},pt^{(1)}_{5}\} $ \\ \hline
\end{tabular}\\
\vspace*{7mm}
{\bf Table 2:} The relation between the bottoms 
and the enhanced  \\
\hspace *{-10.9cm} gauge groups.
\newpage
\begin{tabular}{|l|l|}  \hline
$ \hspace{1.8cm}H $ & \hspace*{1.3cm} Bottom   
\\   \hline \hline
$ SU(2)_{b} $ &  \{$ pt^{(1)}_{2} $\}  \\
$ SU(2)_{c} $ &  \{$ pt^{(1)}_{4} $\}  \\
$ SU(2)_{d} $ &  \{$ pt^{(1)}_{6} $\}  \\
$ SU(2) \times SU(2) $ &  \{$ pt^{(1)}_{2},
pt^{(1)}_{4} $\}  \\
$ (SU(2) \times SU(2))_{b} $ &  \{$ pt^{(1)}_{4},
pt^{(1)}_{6} $\}  \\
$ SU(3) \times SU(2) $ &  \{$ pt^{(1)}_{2},
pt^{(1)}_{4},pt^{(1)}_{5} $\}  \\
$ (SU(3) \times SU(2))_{b} $ &  \{$ pt^{(1)}_{3},
pt^{(1)}_{5} $\}  \\
$ (SU(3) \times SU(2))_{c} $ &  \{$ pt^{(1)}_{5} $\}  \\
$ SO(5) \times SU(2) $ &  \{$ pt^{(1)}_{2},pt^{(1)}_{4},
pt^{(1)}_{6} $\}  \\
$ G_{2} \times SU(2) $ &  \{$ pt^{(1)}_{2},
pt^{(2)}_{4},pt^{(1)}_{5} $\}  \\
$ SU(4) \times SU(2) $ &  \{$ pt^{(1)}_{2},
pt^{(1)}_{4},pt^{(1)}_{5},pt^{(1)}_{6} $\}  \\
$ SO(7) \times SU(2) $ &  \{$ pt^{(1)}_{2},
pt^{(2)}_{4},pt^{(1)}_{5},pt^{(1)}_{6} $\}  \\
$ SO(9) \times SU(2) $ &  \{$ pt^{(1)}_{2},
pt^{(2)}_{4},pt^{(1)}_{5},pt^{(2)}_{6} $\}  \\
$ SU(3)_{b} $ &  \{$ pt^{(1)}_{3} $\}  \\
$ SU(3)_{c} $ &  \{$ pt^{(1)}_{7} $\}  \\
$ SU(3) \times SU(3) $ &  \{$ pt^{(1)}_{3},
pt^{(1)}_{5},pt^{(1)}_{7} $\}  \\ 
$ G_{2} \times SU(3) $ &  \{$ pt^{(1)}_{3},
pt^{(1)}_{5},pt^{(2)}_{7} $\}  \\
$ F_{4} \times SU(2) $ &  \{$ pt^{(1)}_{2},
pt^{(2)}_{4},pt^{(1)}_{5},pt^{(3)}_{6} $\}  
\\   \hline
\end{tabular}\\
\vspace*{7mm}
{\bf Table 3:} The relation between the bottoms
and the enhanced \\ 
\hspace *{-10.9cm} gauge groups.
\end{center}

\end{document}